\documentclass[11pt]{article}

\usepackage{amsmath, amssymb}

\usepackage{graphicx}

\usepackage{geometry}
\usepackage[numbers]{natbib}

\usepackage[hyperfootnotes=false, colorlinks=true, linkcolor=blue, citecolor=blue, urlcolor=blue]{hyperref} 
\usepackage[nameinlink]{cleveref} 

\usepackage{xurl}

\newcommand{\figref}[1]{\cref{#1}}

\title{\Large\itshape Analysis of Regional Disparities of Location of Sports Facilities and Sports Education Using Scientific Satellite Data}

\author{Shoichi Otomo}

\date{July 2026}

\begin{document}

\maketitle

\begin{abstract}

This manuscript is an English translation and extended version of a paper originally published in Japanese~\cite{otomo3}.

With the advancement of information and communication technology (ICT) and data analysis techniques, handling massive datasets (big data) has become feasible in spatial information science. Consequently, infrastructure is being established to generate new utility value from spatial data. Furthermore, nationwide initiatives are actively promoting the use of open data—public information provided under terms allowing secondary usage.

However, in developing regions and specific municipalities, local statistical data remain scarce, unreliable, or difficult to acquire. Additionally, long-standing concerns exist regarding the inherent limitations of standard macroeconomic indicators when performing cross-regional or international comparisons.

Meanwhile, high-frequency satellite data has become widely accessible. Among various observational products, nightlight data has attracted significant attention due to its versatile applications. 

Generally, nightlight radiance correlates with urbanization. Prior work indicates that areas with larger populations and higher commercial development display higher radiance, whereas public infrastructure—such as sanitation systems and healthcare facilities—shows little correlation with nightlight intensity~\cite{otomo}. Building on these findings, this paper analyzes the relationship between nightlight data and the spatial distribution of sports facilities, the evolution of the fitness industry, and regional environmental disparities in sports access. The analysis confirms that the spatial distribution patterns of sports facilities differ significantly between the public and private sectors.

\end{abstract}

%%%%%%%キーワードの設定

\section{Introduction}

\subsection{Literature Review}
The widespread application of Earth observation data in socio-economic research stems from recent breakthroughs in big data processing and public archive access. As Donaldson et al. (2016) \cite{donaldson} outlined, the availability of high-resolution, petabyte-scale imagery has transformed traditional spatial analysis across economics, urban planning, and geography.

Nighttime light (NTL) remote sensing, in particular, has proven highly adaptable. Henderson et al. (2012) \cite{henderson} established that NTL radiance serves as a robust proxy for economic output during sudden socio-economic shocks, offering a unified observational framework that circumvents measurement noise, reporting bias, and currency conversion issues present in standard GDP figures. In regional development, Ichinose et al. (2002) \cite{ichinose} showcased the efficacy of GIS-based spatial aggregation for quantifying urban expansion patterns across Asian cities when municipal statistics are unavailable. Furthermore, Kurata (2017) \cite{kurata} confirmed that NTL intensity correlates strongly with local demographic patterns, infrastructure maturity, and human development indicators in low-income districts.

Expanding on spatial dynamics, Li et al. (2019) \cite{changli} utilized NTL data to evaluate spatial economic spillovers and fill statistical gaps across Chinese provinces. In the domain of seasonal human activity, Chalkias et al. (2019) \cite{chalkias} applied linear and geographically weighted regression (GWR) models to show that satellite-observed light intensity closely tracks seasonal tourism concentration across the European Union. In energy management, Derek and Moncef (2018) \cite{FEHRER2018252} extended NTL analytics to downscale spatial energy consumption modeling to building-level scales.

Despite these advances in spatial big data, quantitative spatial analytics remain underutilized in sports facility management. Historically, as Karube (2002) \cite{karube2002} noted, research on sports facility location has focused largely on qualitative conceptualizations or localized case studies, lacking broad quantitative frameworks suitable for evidence-based policy making and urban facility planning.

\subsection{Nightlight Data Used}
This study uses nighttime light data from the Defense Meteorological Satellite Program (DMSP) operational satellites, distributed by the National Oceanic and Atmospheric Administration (NOAA) as annual composites and radiance-calibrated GeoTIFF products \cite{geotiff}. According to NOAA specifications \cite{noaa3}, the grid spans globally from $-180^\circ$ to $180^\circ$ longitude and $-65^\circ$ to $75^\circ$ latitude, with a spatial resolution of 30 arc-seconds per cell. A inherent constraint of this sensor resolution is that closely co-located infrastructure—such as multi-sport complexes containing adjacent athletic fields, ballparks, and swimming facilities—cannot be individually isolated based solely on NTL radiance.

\subsection{Significance and Novelty of This Study}
To overcome the spatial and sample limitations of prior work—such as Karube (2002) \cite{karube2002}, which examined 155 facilities within a single prefecture—this study conducts a comprehensive nationwide analysis covering 10,683 sports facilities across Japan. Utilizing Geographic Information System (GIS) platforms including ArcGIS \cite{arcgis} and QGIS \cite{qgis}, I perform proximity analysis at a 100-meter resolution, providing significantly higher precision than the 1-km grid intervals used in legacy literature.

By integrating high-resolution proximity calculations with satellite NTL radiance, this study reveals distinct spatial distribution patterns: private sports facilities exhibit strong spatial clustering in high-radiance urban cores, whereas public facilities display a more dispersed distribution across lower-radiance peripheral zones. These empirical insights and methodological advances represent the principal contributions of this paper.

\section{Data Analysis of the Sports Industry}
\subsection{Trends in the Market Size of the Fitness Industry}

Data relevant to the fitness industry from the Current Survey of Selected Service Industries, available on e-Stat \cite{estat}, provides monthly data on the number of members, sales, and other metrics starting from January 2001. These trends are visualized in \figref{fig:fitness_sales}, \figref{fig:fitness_user}, and \figref{fig:fitness_labor} below.

All monetary values originally in JPY are converted to USD (and EUR) based on the Bank of Japan exchange rates as of August 3, 2026 ($1~\text{USD} = 156.50~\text{JPY}$ and $1~\text{EUR} = 180.69~\text{JPY}$) \cite{jpbank}. The resulting monetary values are expressed in thousand USD [and EUR] in the figures.

%水泳pic1
%%%%%%%%%%はじまり%%%%%%%%%%%%%%
\begin{figure}
\begin{center}
\includegraphics[width=130mm]{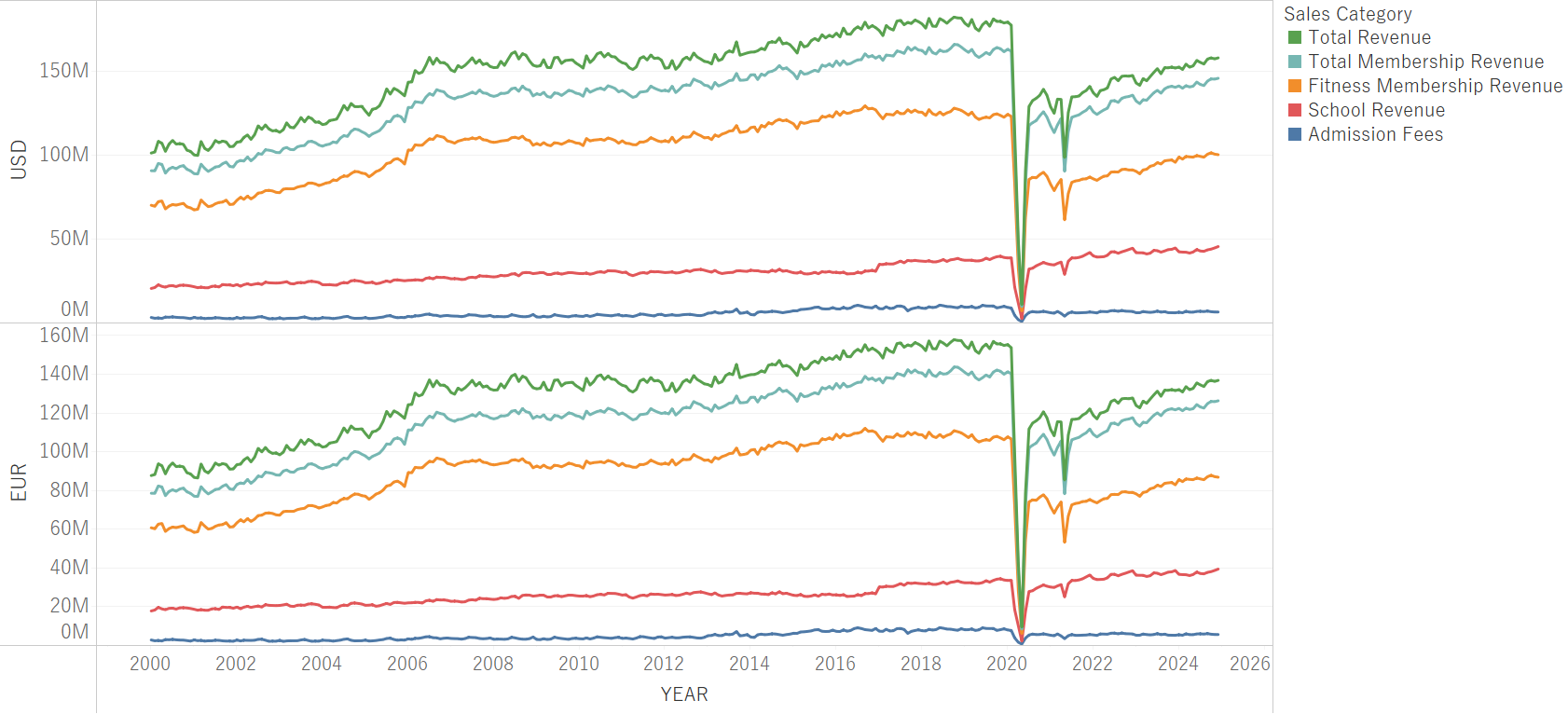}
\end{center}
\caption{Trends in Net Sales, Membership Fee Revenue}
\label{fig:fitness_sales}
\end{figure}
%%%%%%%%%%おわり%%%%%%%%%%%%%%%%

From the start of the survey in 2001 until the onset of the COVID-19 pandemic in 2019, the industry experienced continuous growth. Monthly user numbers escalated from approximately 9 million in January 2001 to a peak of over 22 million in July 2019. Concurrently, monthly sales increased from roughly 102,236 thousand USD (88,549 thousand EUR) to over 178,914 thousand USD (154,961 thousand EUR). Reflecting this expansion, the number of employees also grew from about 20,000 to a peak of over 45,000. These indicators demonstrate that the fitness industry as a whole was undergoing a period of robust expansion.
%水泳pic2
%%%%%%%%%%はじまり%%%%%%%%%%%%%%
\begin{figure}
\begin{center}
\includegraphics[width=130mm]{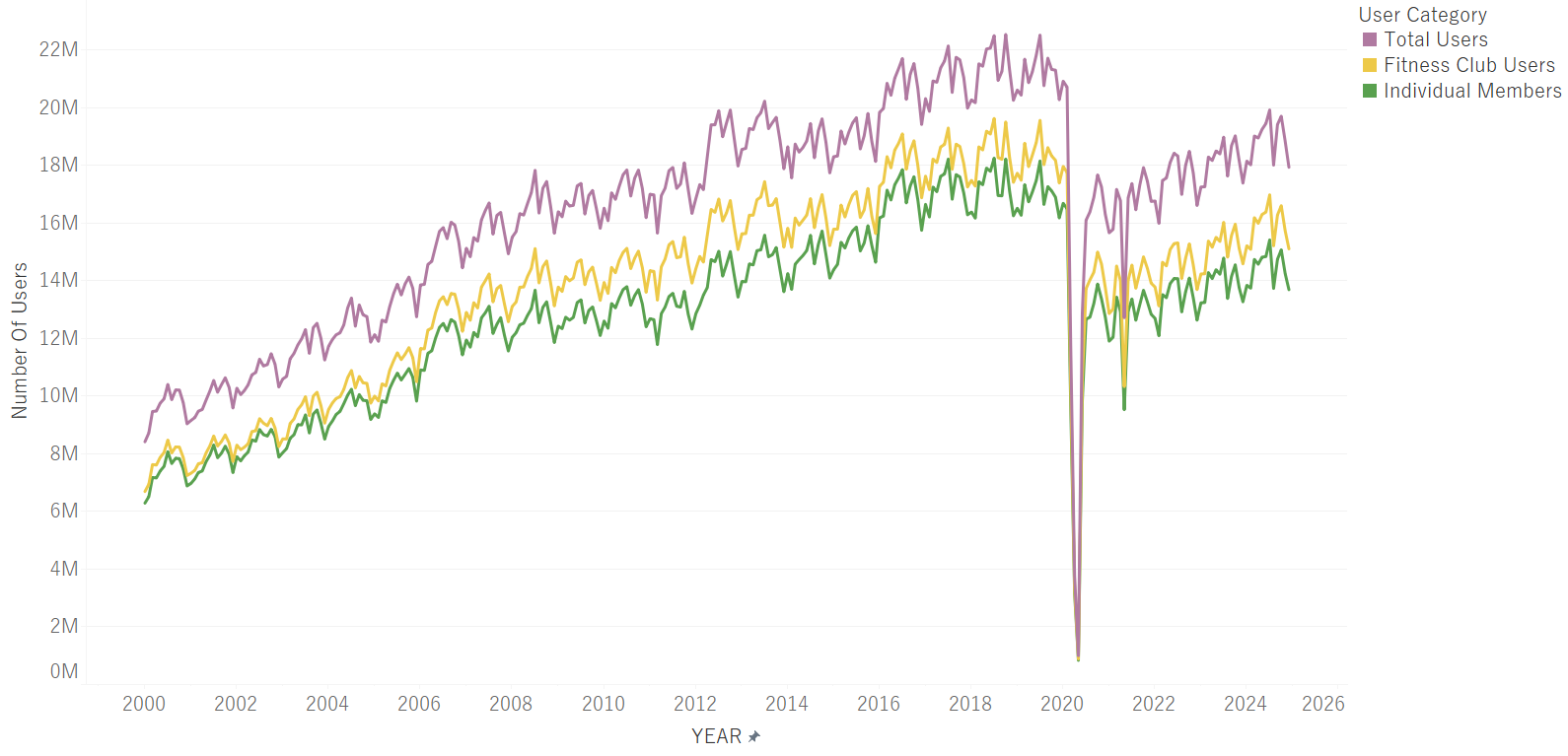}
\end{center}
\caption{Trends in Number of Users, Members}
\label{fig:fitness_user}
\end{figure}
%%%%%%%%%%おわり%%%%%%%%%%%%%%%%
%水泳pic3
%%%%%%%%%%はじまり%%%%%%%%%%%%%%
\begin{figure}
\begin{center}
\includegraphics[width=130mm]{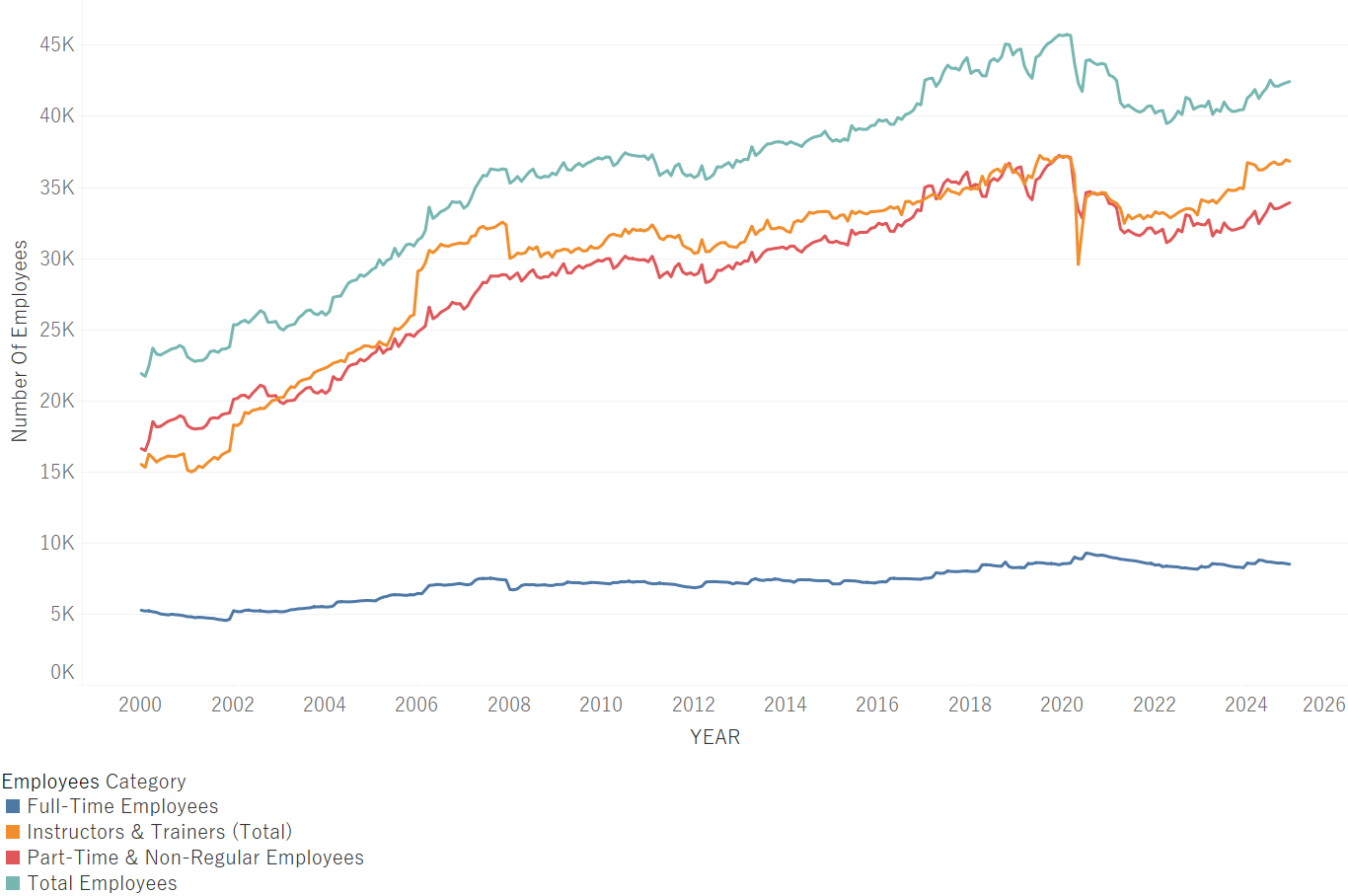}
\end{center}
\caption{Trends in Number of employees}
\label{fig:fitness_labor}
\end{figure}
%%%%%%%%%%おわり%%%%%%%%%%%%%%%%

\subsection{Locations of Sports Facilities}
\begin{figure}[htbp]
  \centering
  \includegraphics[width=130mm]{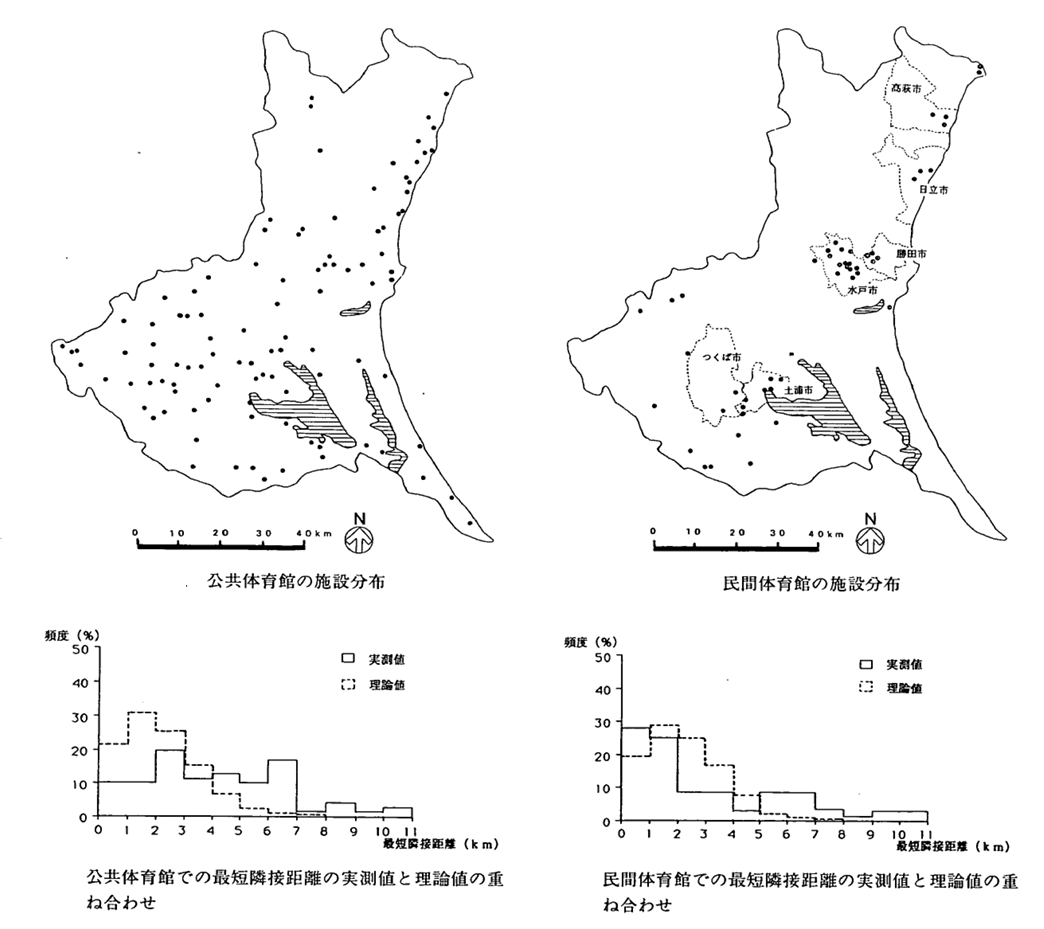}
  \caption[Spatial Distribution of Sports Facilities]{Spatial Distribution of Sports Facilities (Public vs. Private)\footnotemark}
  \label{fig:fitness_facets}
\end{figure}

Regarding the spatial distribution of sports facilities, according to the Statistics of Sports \cite{sports_stats} and Karube (2002) \cite{karube2002} (as shown in \figref{fig:fitness_facets}), public facilities are evenly distributed across individual municipalities. In contrast, private facilities are clustered in relatively populous cities, such as Mito, Hitachi, and Tsukuba.

%fitness pic4
%%%%%%%%%% BEGINNING %%%%%%%%%%%%%%
\footnotetext{Partially modified by the author based on figures in \cite{sports_stats} and \cite{karube2002}. Original text in the figure is in Japanese.}
%%%%%%%%%% END %%%%%%%%%%%%%%%%

Following this precedent, this paper calculated the nearest-neighbor distances for the nationwide locations of sports facilities for the fiscal year 2021. The results are presented in \figref{fig:fitness_pub} and \figref{fig:fitness_prv} below. As is evident from these figures, consistent with prior research, private facilities across Japan exhibit shorter nearest-neighbor distances compared to public facilities. This indicates that private facilities are densely concentrated in specific locations.

According to Otomo \cite{otomo}, commercial stores, particularly in the service and retail industries, are known to cluster in areas with high nightlight intensity. Conversely, highly public services, such as healthcare, exhibit a relatively low correlation with nightlights. Given these characteristics, this study examines where sports facilities are distributed or clustered, and the extent to which these spatial patterns differ between public and private sectors.

%fitness pic5
%%%%%%%%%%はじまり%%%%%%%%%%%%%%
\begin{figure}
\begin{center}
\includegraphics[width=130mm]{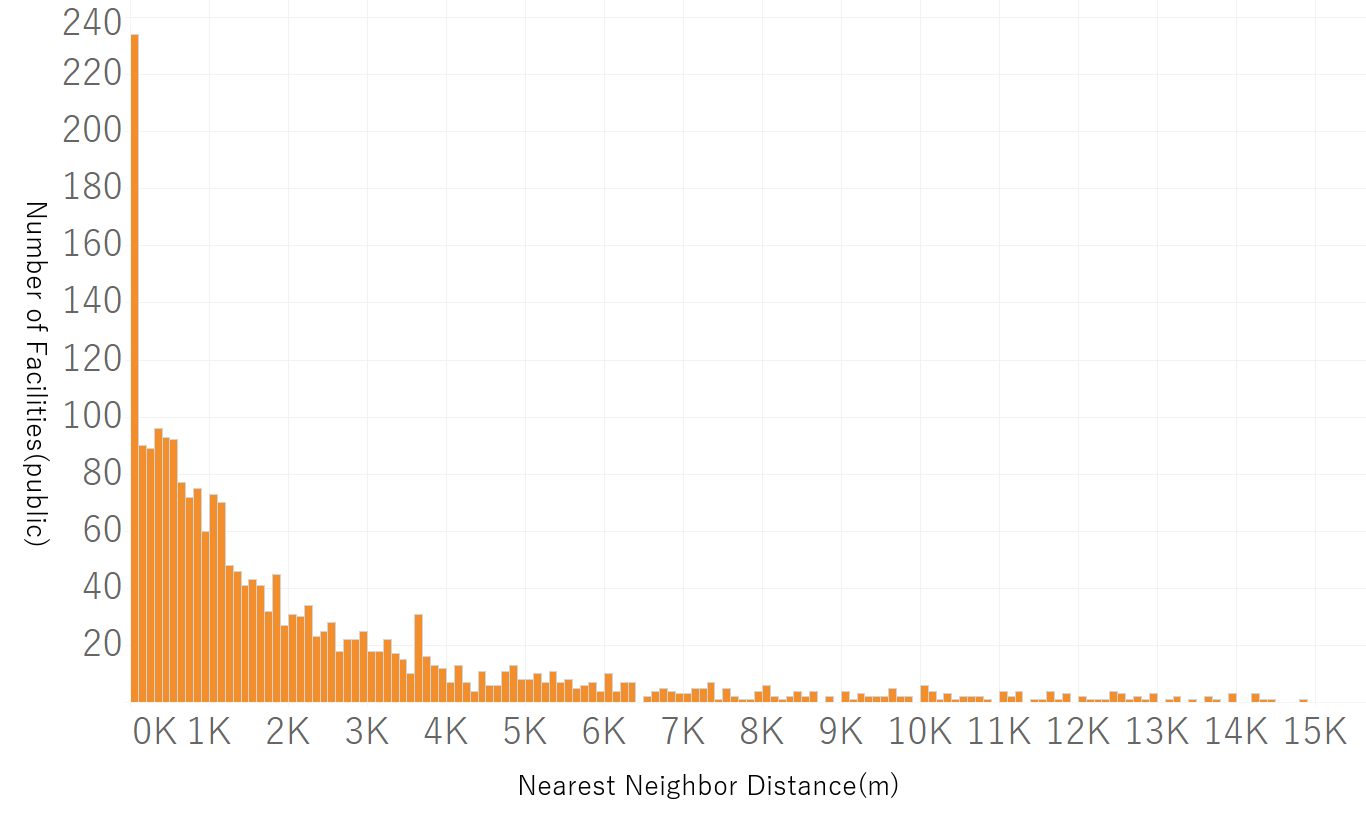}
\end{center}
\caption{Shortest distance to the nearest public sports facility}
\label{fig:fitness_pub}
\end{figure}
%%%%%%%%%%おわり%%%%%%%%%%%%%%%%
%fitness pic6
%%%%%%%%%%はじまり%%%%%%%%%%%%%%
\begin{figure}
\begin{center}
\includegraphics[width=130mm]{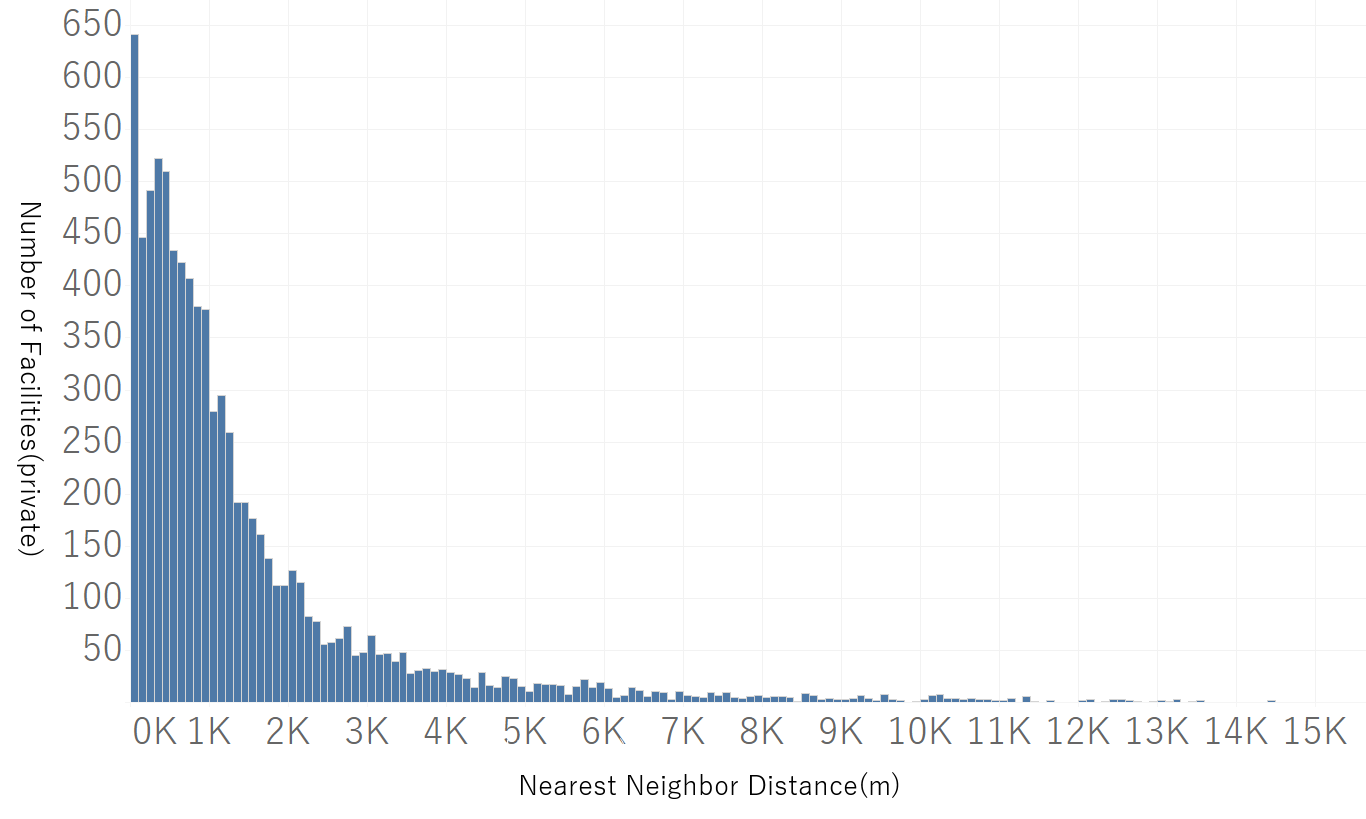}
\end{center}
\caption{Shortest distance to the nearest private sports facility}
\label{fig:fitness_prv}
\end{figure}
%%%%%%%%%%おわり%%%%%%%%%%%%%%%%

%水泳pic7
%%%%%%%%%%はじまり%%%%%%%%%%%%%%
\begin{figure}
\begin{center}
\includegraphics[width=130mm]{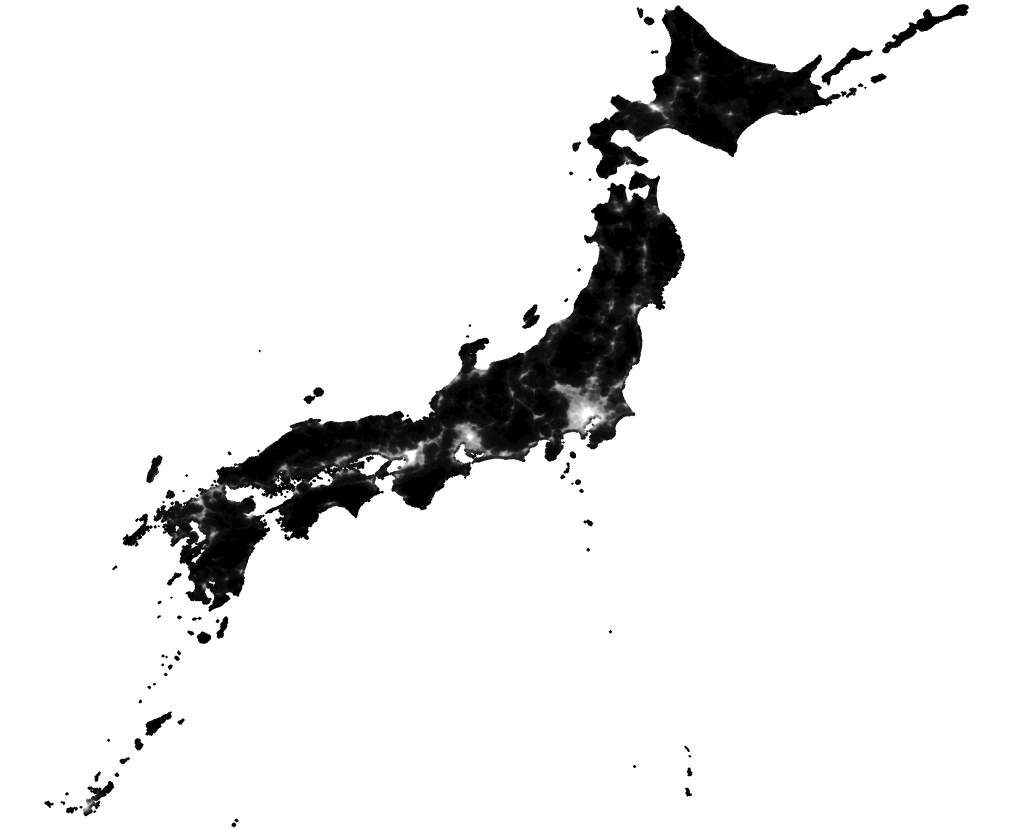}
\end{center}
\caption[jpn]{Nighttime light brightness in Japan\footnotemark}
%footermark[3]は論文内参照のためimgをスキップした
\label{fig:night_light_jpn2}
\end{figure}
\footnotetext{Origin-Destination NOAA, \cite{noaa1}, Data values range from 0 to 63.}
%%%%%%%%%%おわり%%%%%%%%%%%%%%%%

\section{Analysis and Discussion Using Nightlight Data}
\subsection{Distribution of Sports Facilities Based on Nightlight Intensity}

To briefly describe the characteristics of nightlights across Japan, referencing Otomo \cite{otomo}\cite{otomo2} (as shown in \figref{fig:night_light_jpn2}), it is evident that metropolitan centers are bright throughout. In contrast, the nightlight characteristics of less urbanized municipalities do not simply mean that their minimum values are zero. While commercial areas, such as those around train stations, are bright, many locations within the same municipality record a minimum value of zero; consequently, the variance tends to be large.

To investigate the locational distribution of sports facilities, data were extracted using NTT's online telephone directory, "i-Town Page" \cite{itp}. Subsequently, to append latitude and longitude coordinates to these facility data, I utilized the geocoding service provided by the Center for Spatial Information Science at the University of Tokyo \cite{csis}.

\begin{figure}[htbp]
  \centering % \begin{center} よりも余白が綺麗に制御できる
  
  \includegraphics[width=130mm]{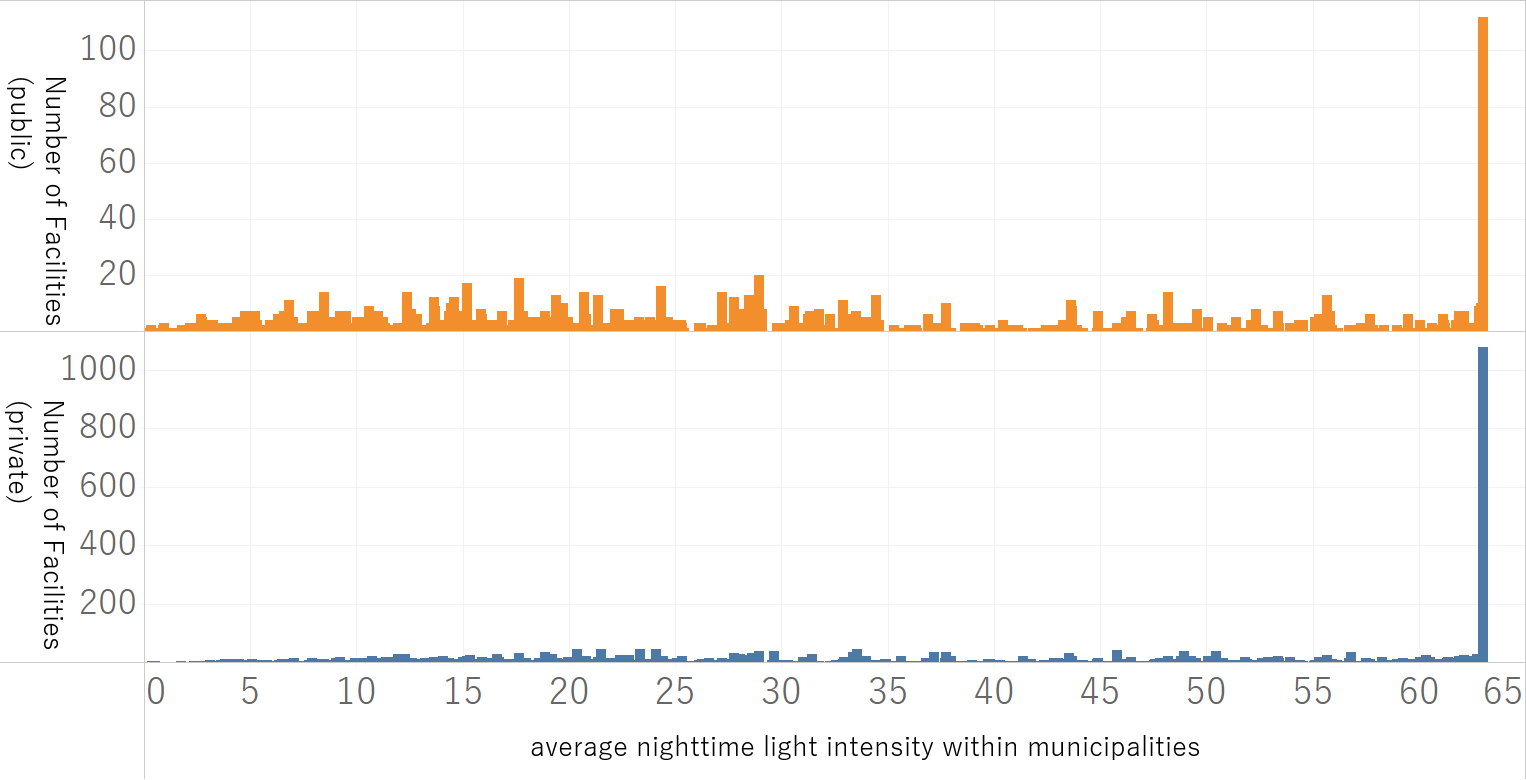}
  \caption{Number of sports facilities by average nighttime light intensity within municipalities}
  \label{fig:fitness_night_light}

  \vspace{10mm} % 必要に応じて画像間の垂直方向の隙間を調整

  \includegraphics[width=130mm]{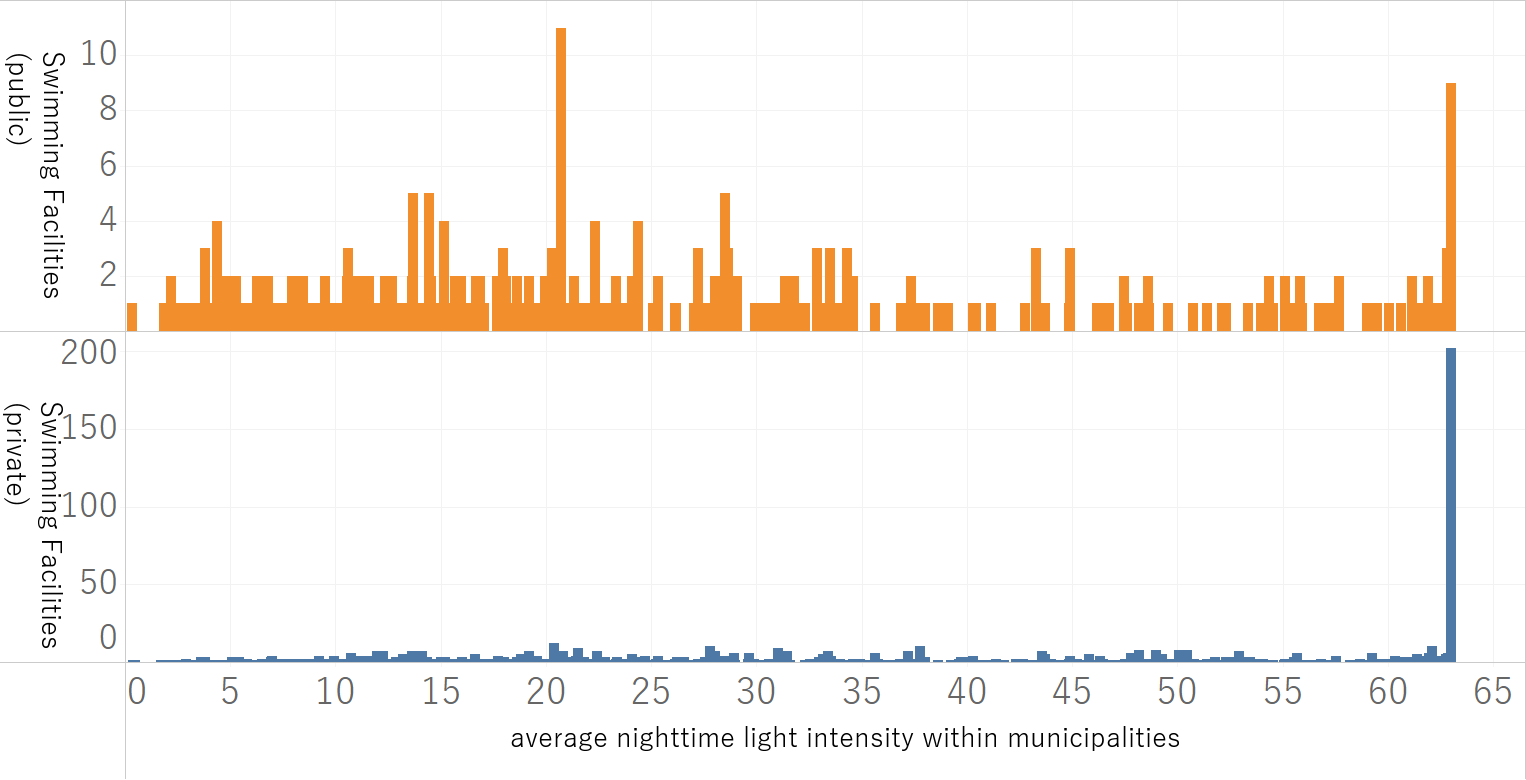}
  \caption{Number of swimming facilities by average nighttime light intensity within the municipality}
  \label{fig:swim_night_light}
\end{figure}

First, \figref{fig:fitness_night_light} and \figref{fig:swim_night_light} illustrate the distribution of the number of facilities relative to the mean nightlight intensity within each municipality for both sports and swimming facilities, respectively. These figures indicate that private facilities—both sports and swimming—are densely clustered in areas with high nightlight intensity; that is, relatively bright areas such as commercially developed station fronts, downtown districts, or residential zones.
On the other hand, although public sports facilities are also frequently located in areas with high nightlight intensity, their degree of clustering is lower than that of private facilities.
This spatial pattern suggests that when establishing public facilities, local governments provide infrastructure and related maintenance operations to regional areas by taking factors other than user convenience into account. As a result, public facilities are distributed more evenly across areas beyond those with high nightlight intensity.

Next, according to the Parent-Child Survey on Children's Life and Learning \cite{ut_benesse}, swimming ranked first among extracurricular activities for both lower- and upper-grade elementary school boys. Therefore, this section presents the measured nearest-neighbor distances for swimming facilities—specifically focusing on dedicated swimming facilities or complex facilities equipped with swimming pools—in the same manner as the general sports facilities.

\begin{figure}
  \centering

  % --- 1枚目 (pic10) ---
  \begin{minipage}{0.48\textwidth}
    \centering
    \includegraphics[width=\textwidth]{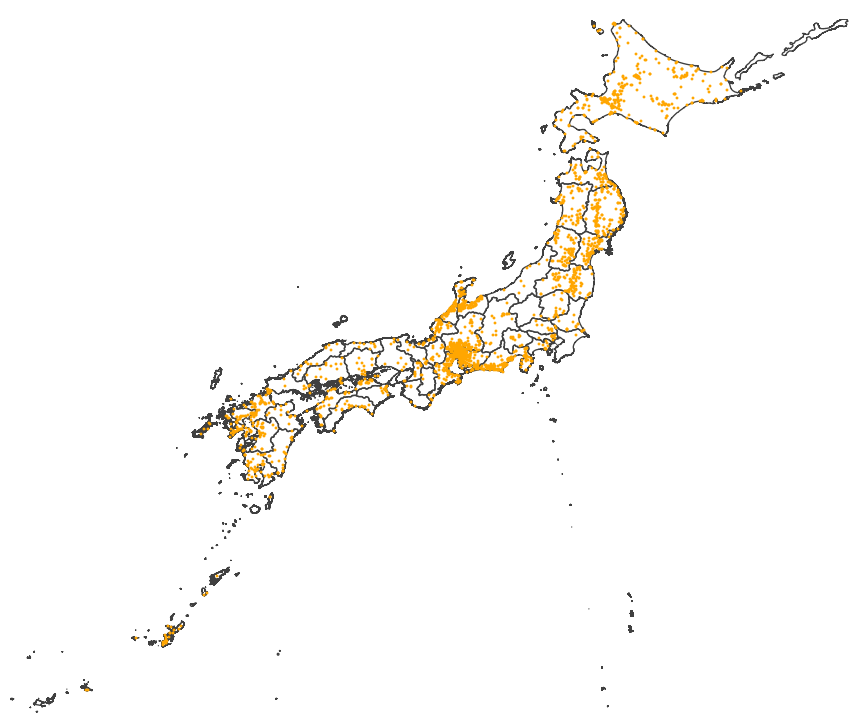}
    \caption{Distribution of public sports facilities in Japan}
    \label{fig:fitness_jpn_pub}
  \end{minipage}
  \hspace{2mm} %左右の隙間を10mmに固定
  % --- 2枚目 (pic11) ---
  \begin{minipage}{0.48\textwidth}
    \centering
    \includegraphics[width=\textwidth]{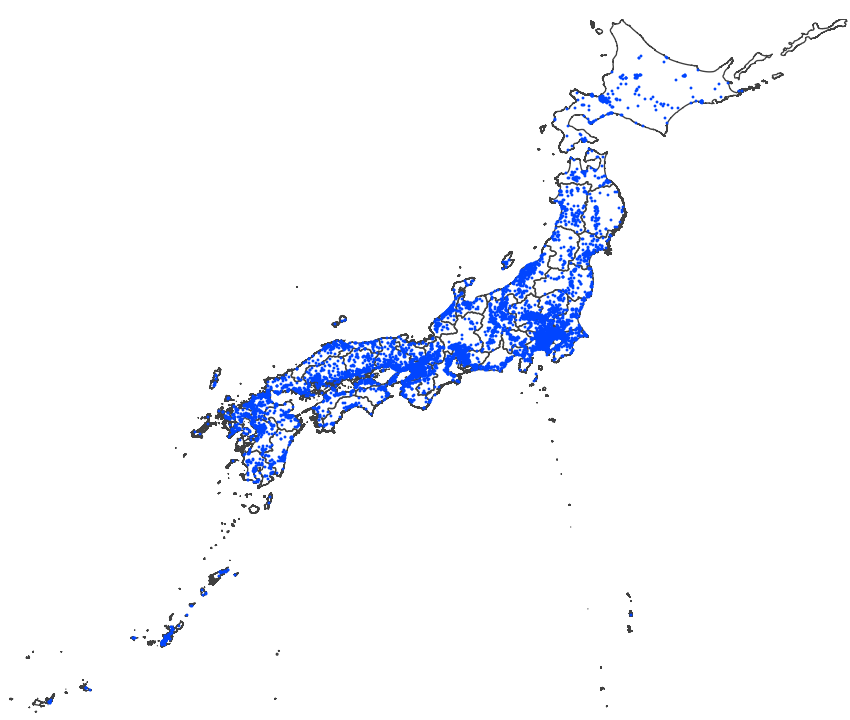}
    \caption{Distribution of private sports facilities in Japan}
    \label{fig:fitness_jpn_prv}
  \end{minipage}

  \vspace{20mm} % 上下の行の間の余白

  % --- 3枚目 (pic12) ---
  \begin{minipage}{0.48\textwidth}
    \centering
    \includegraphics[width=\textwidth]{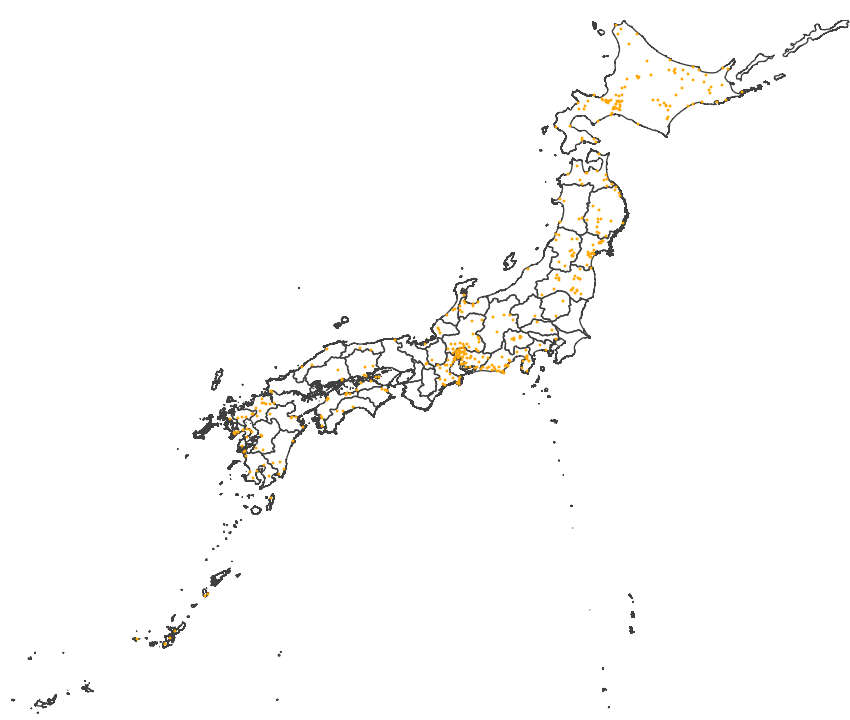}
    \caption{Distribution of public swimming facilities in Japan}
    \label{fig:jpn_pub_swim2.png}
  \end{minipage}
  \hspace{2mm} %左右の隙間を10mmに固定
  % --- 4枚目 (pic13) ---
  \begin{minipage}{0.48\textwidth}
    \centering
    \includegraphics[width=\textwidth]{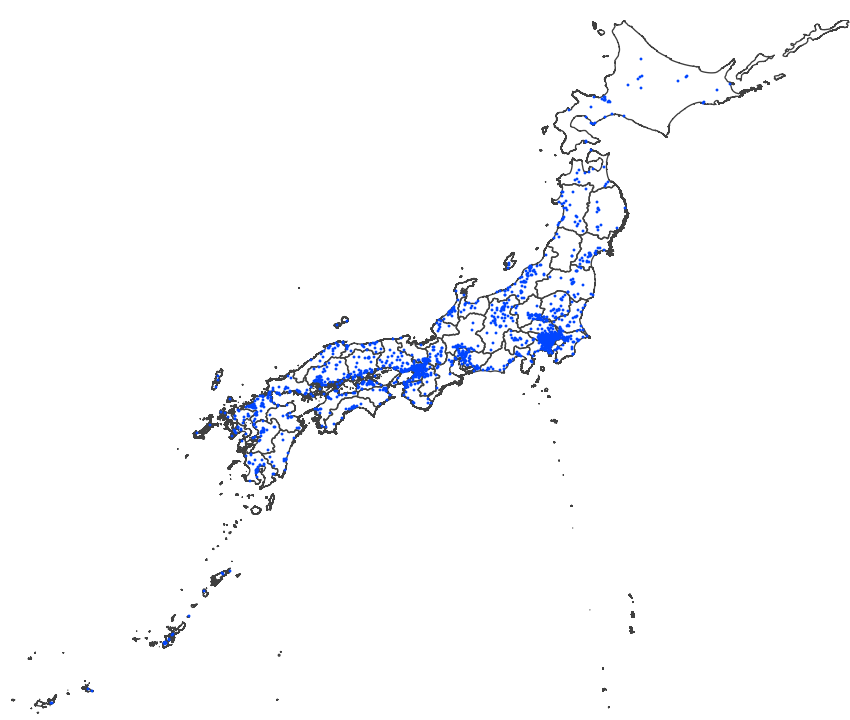}
    \caption{Distribution of private swimming facilities in Japan}
    \label{fig:jpn_prv_swim2.png}
  \end{minipage}
\end{figure}

As a result, \figref{fig:fitness_night_light} and \figref{fig:swim_night_light} reveal that, compared to general sports facilities, swimming facilities exhibit a more pronounced tendency for private sectors to be concentrated in areas with higher nightlight intensity. This geographical distribution of facilities is visually mapped in \figref{fig:fitness_jpn_pub}, \figref{fig:fitness_jpn_prv}, \figref{fig:jpn_pub_swim2.png}, and \figref{fig:jpn_prv_swim2.png}\footnotemark. \footnotetext{Partially modified from \cite{otomo} and \cite{otomo2}.}
In the following chapter, I examine the number of participants in national tournaments by prefecture, which is assumed to be influenced by the location of these swimming facilities.

%%%%%%%%%%はじまり%%%%%%%%%%%%%%
%%%%%%%%%%おわり%%%%%%%%%%%%%%%%

%%%%%%%%%%%体育施設・水泳施設地図はじまり%%%%%%%%%%%
% 4枚の画像を1ページに集約
%%%%%%%%%%%体育施設・水泳施設地図おわり%%%%%%%%%%%

%%%第3章からでclearpageブレイクする．newpageはページが切り替わらない
\clearpage

\section{Data Analysis of Swimming Tournaments}

\subsection{National Junior High School Swimming Championship}

\begin{figure}
\begin{center}
\includegraphics[width=130mm]{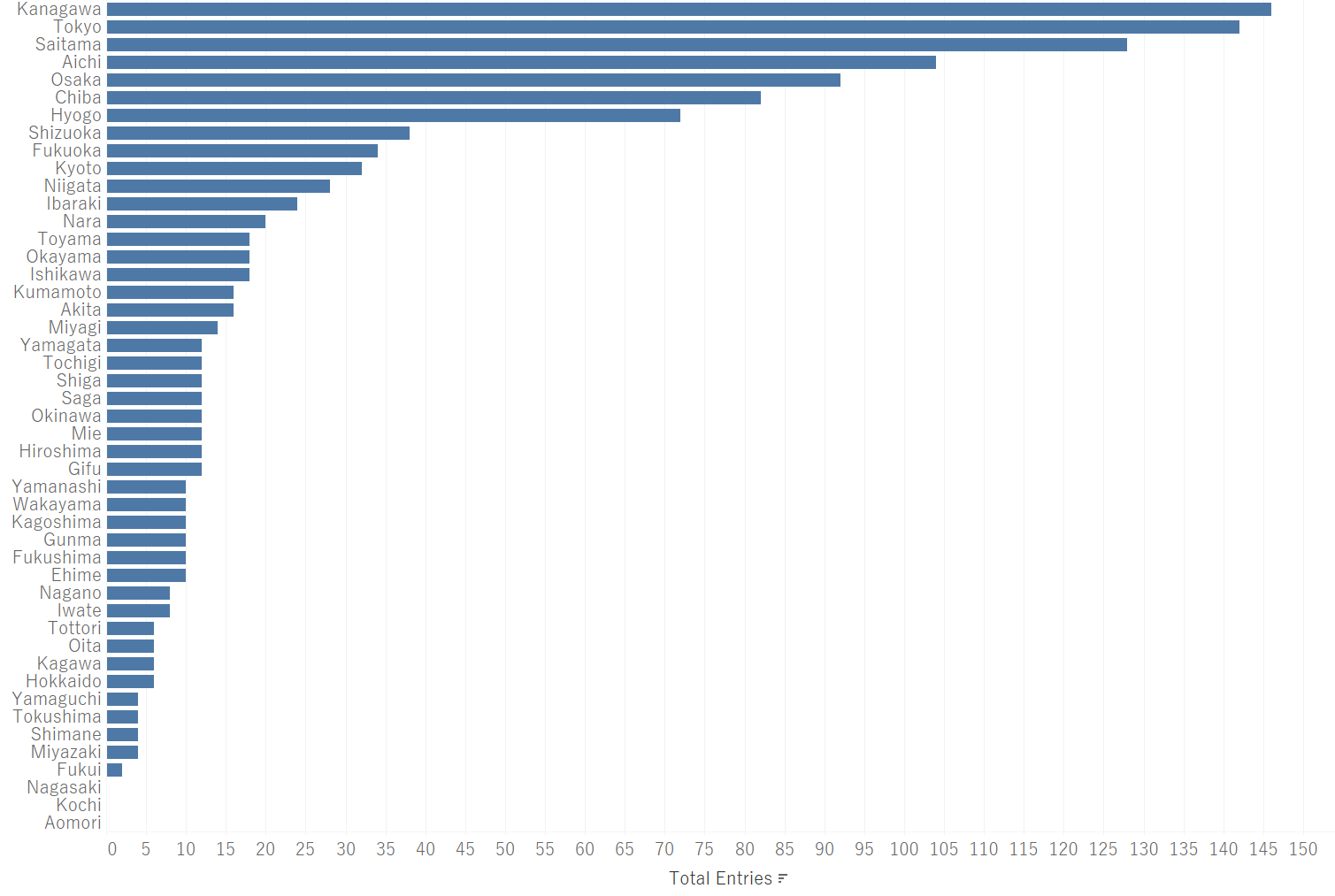}
\end{center}
\caption{2021 National Junior High School Swimming Championships: Number of Appearances by Prefecture}
\label{fig:swim_juniorhigh2021}
\end{figure}

Based on the data publicly released by SEIKO \cite{seiko}, an examination of the tournament results for the 2021 National Junior High School Swimming Championship (\figref{fig:swim_juniorhigh2021}) reveals that the prefectures with the highest number of tournament appearances are, in descending order, Kanagawa, Tokyo, Saitama, Aichi, Osaka, and Chiba. As is evident from \figref{fig:night_light_jpn2}, all of these prefectures are metropolitan areas characterized by high nightlight intensity and a dense concentration of private swimming facilities.

%水泳pic14
%%%%%%%%%%はじまり%%%%%%%%%%%%%%
%%%%%%%%%%おわり%%%%%%%%%%%%%%%%
%水泳pic15
%%%%%%%%%%はじまり%%%%%%%%%%%%%%
\begin{figure}
\begin{center}
\includegraphics[width=130mm]{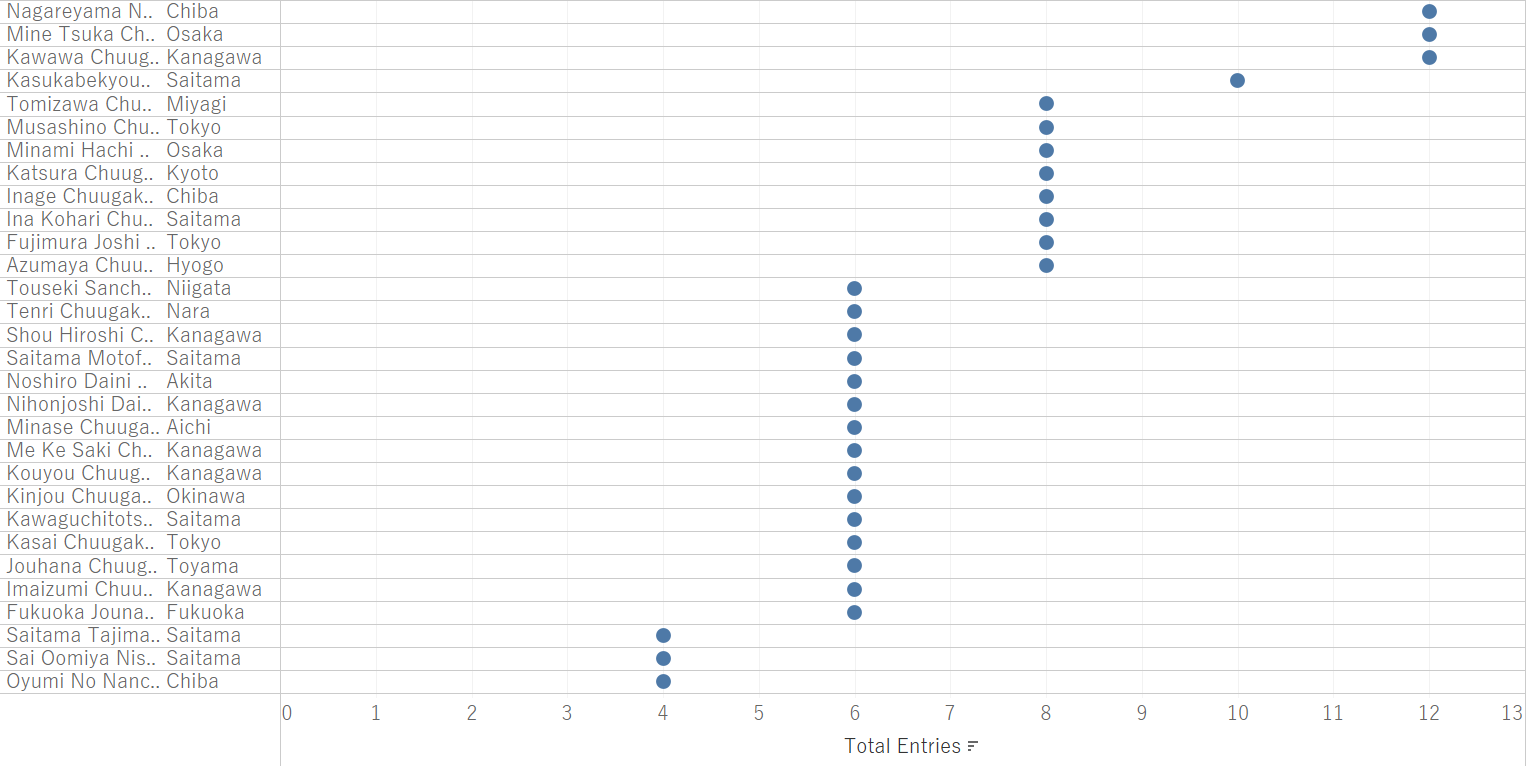}
\end{center}
\caption{2021 National Junior High School Swimming Championships: Number of Appearances by School}
\label{fig:swim_juniorhigh2021school}
\end{figure}
%%%%%%%%%%おわり%%%%%%%%%%%%%%%%
%
%水泳pic16
%%%%%%%%%%はじまり%%%%%%%%%%%%%%
\begin{figure}
\begin{center}
\includegraphics[width=130mm]{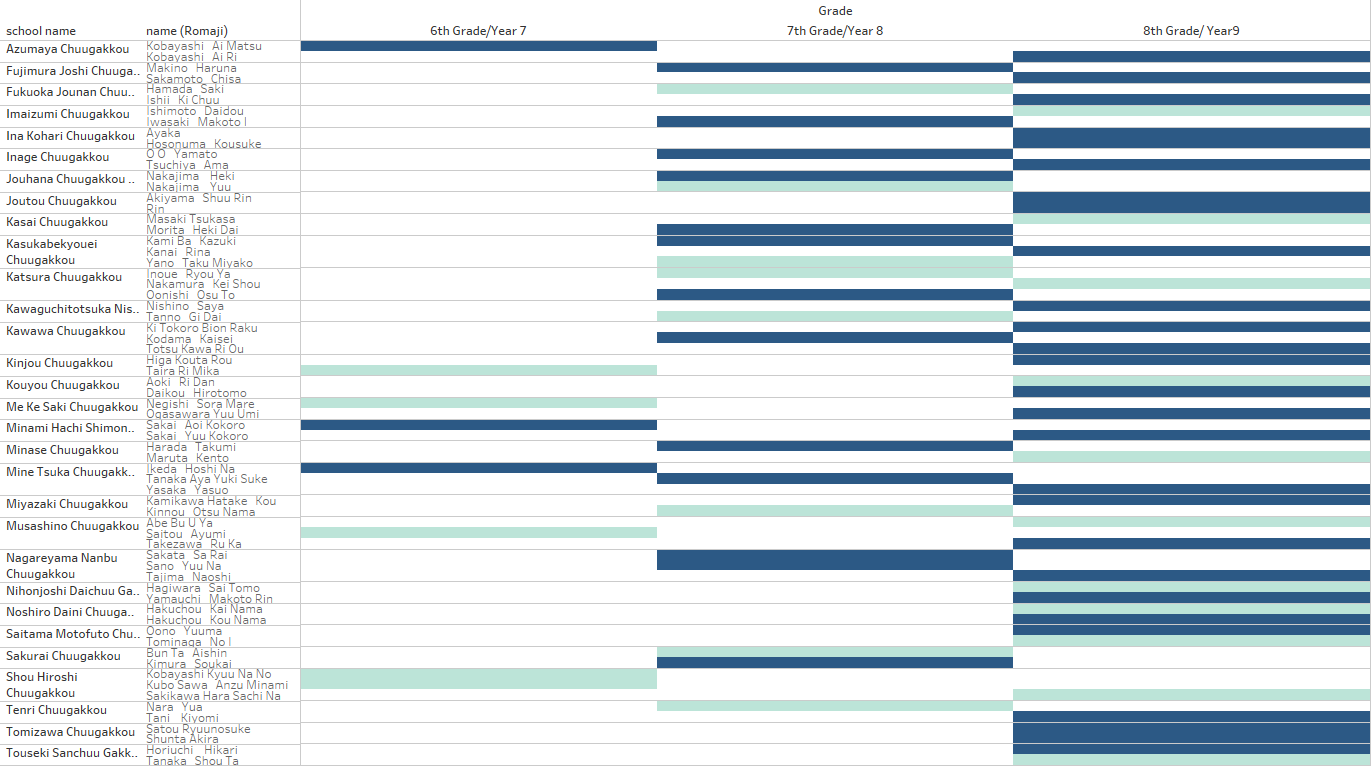}
\end{center}
\caption{2021 National Junior High School Swimming Championships: Number of Appearances by School Year and Athlete}
\label{fig:swim_juniorhigh2021school_grade}
\end{figure}
%%%%%%%%%%おわり%%%%%%%%%%%%%%%%

At the junior high school level, sports recommendations (athletic recruitment) generally do not exist. Nevertheless, \figref{fig:swim_juniorhigh2021school} clearly indicates that certain junior high schools account for a disproportionately high number of tournament appearances.

Furthermore, up to the junior high school age, swimmers often do not specialize in a single stroke; instead, talented athletes are typically trained across all individual medley disciplines (freestyle, breaststroke, butterfly, and backstroke). Consequently, as shown in \figref{fig:swim_juniorhigh2021school_grade}, it is common for a single athlete to compete in multiple events. For this reason, \figref{fig:swim_juniorhigh2021school} and \figref{fig:swim_juniorhigh2021school_grade} define the metrics in terms of the number of "appearances" rather than the number of unique participants.

This raises the question of whether these schools should be classified as so-called "powerhouse schools" (dominant athletic programs).

While it is plausible that some private junior high schools coordinate with their affiliated high schools or universities to foster athlete development, it remains highly questionable whether public junior high schools can establish similar pipelines.

In either case, because these schools are geographically situated near metropolitan centers, an alternative hypothesis can be considered for talented athletes attending public junior high schools: these athletes train at private swimming facilities, and because they attend these specific facilities, they naturally enroll in junior high schools within the corresponding local school districts.

\subsection{National High School Swimming Championship}

\begin{figure}
\begin{center}
\includegraphics[width=130mm]{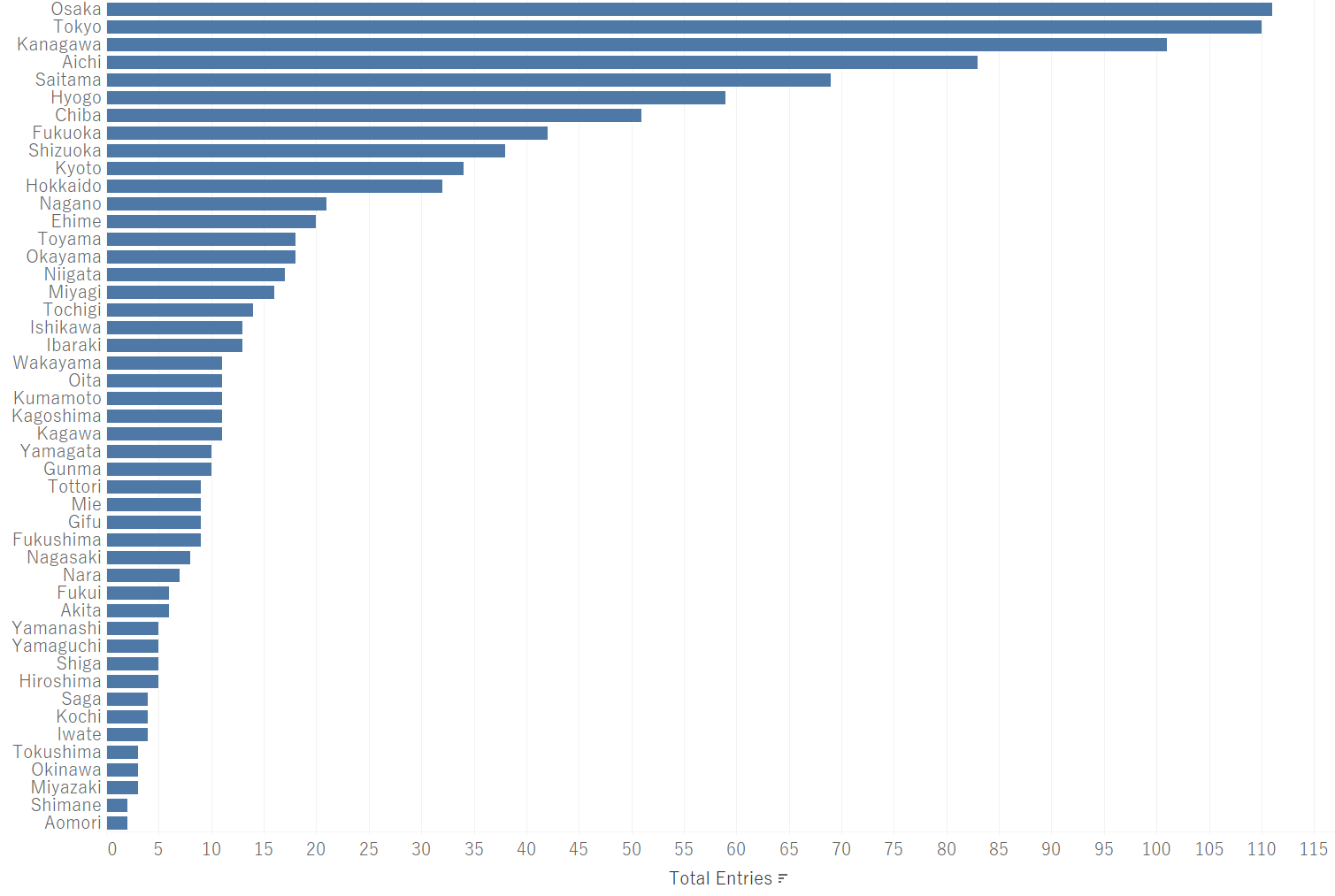}
\end{center}
\caption{2021 National High School Swimming Championships: Number of Appearances by Prefecture}
\label{fig:swim_high2021}
\end{figure}

An examination of the tournament results for the 2021 National High School Swimming Championship (\figref{fig:swim_high2021}) reveals that the prefectures with the highest number of tournament appearances are, in descending order, Osaka, Tokyo, Kanagawa, Aichi, and Saitama. Although the exact ranking varies slightly, these results are nearly identical to the top prefectures observed in the National Junior High School Swimming Championship.

%水泳pic17
%%%%%%%%%%はじまり%%%%%%%%%%%%%%
%%%%%%%%%%おわり%%%%%%%%%%%%%%%%
%水泳pic18
%%%%%%%%%%はじまり%%%%%%%%%%%%%%
\begin{figure}
\begin{center}
\includegraphics[width=130mm]{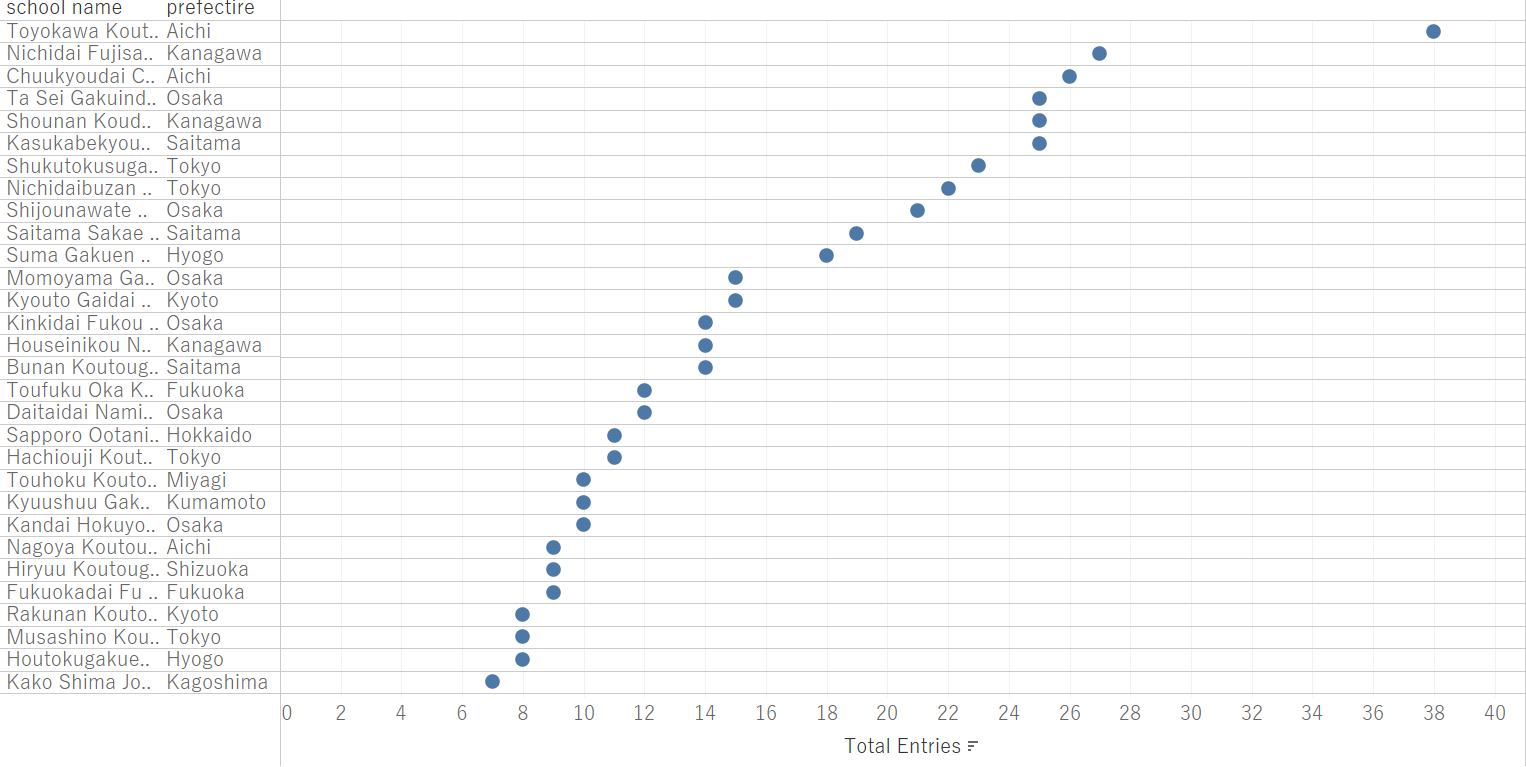}
\end{center}
\caption{2021 National High School Swimming Championships: Number of Appearances by High School}
\label{fig:swim_high2021school}
\end{figure}
%%%%%%%%%%おわり%%%%%%%%%%%%%%%%

%水泳pic18
%%%%%%%%%%はじまり%%%%%%%%%%%%%%
\begin{figure}
\begin{center}
\vspace*{-15mm} % ここで数値を調整（マイナス値で上にズレる）
\includegraphics[width=130mm]{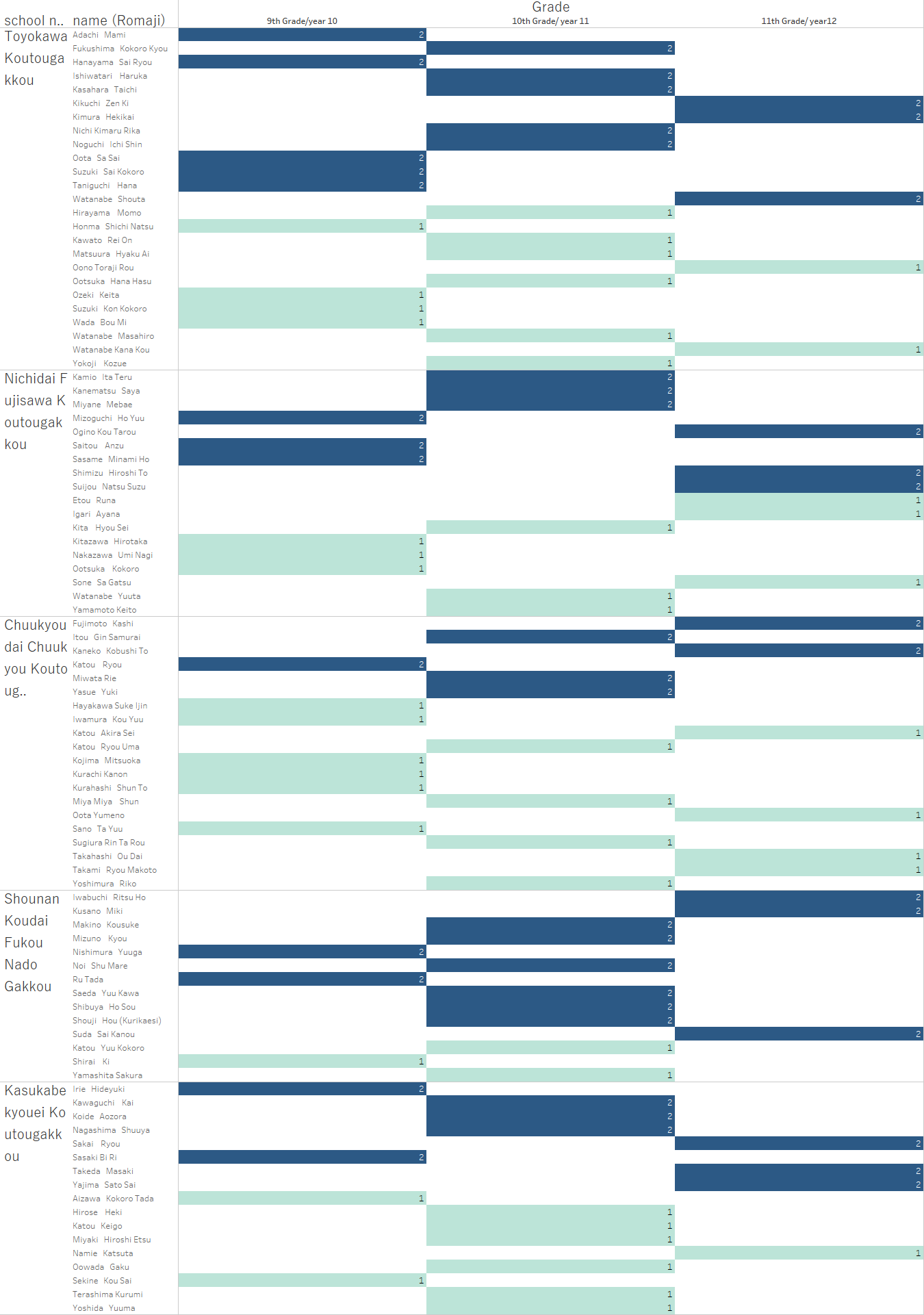}
\end{center}
\caption{Number of Appearances by High School, Grade, and Athlete at the National High School Swimming Championships}
\label{fig:swim_high2021school_grade}
\end{figure}
%%%%%%%%%%おわり%%%%%%%%%%%%%%%%

Unlike junior high schools, high schools generally utilize athletic recruitment (sports recommendations). Presumably reflecting this institutional difference, a classification of schools into public and private sectors based on the MEXT School Codes \cite{monka} reveals that among the top 20 junior high schools (which expand to 27 schools when including ties), only 5 are private, accounting for less than 20\% (\figref{fig:swim_juniorhigh2021school}). 

In contrast, at the high school level (\figref{fig:swim_high2021school}), all of the top 20 schools are private. Furthermore, within the top 30 high schools (34 schools including ties), only a single public school is present, demonstrating a substantial increase in the proportion of private institutions.

As is evident from \figref{fig:swim_high2021school_grade}, while some athletes still compete in multiple events, the total number of participants from a single school increases significantly compared to junior high schools. This shift is likely attributable to the fact that swimmers typically begin to specialize in specific strokes around high school age.

Moreover, to gain admission to elite high school swimming programs via athletic recruitment, athletes must achieve outstanding results during junior high school. Based on the analysis presented thus far, achieving such performance at the junior high school level implies that these athletes train at private swimming facilities. Extending this timeline, to secure top marks at the National Junior High School Swimming Championship, it is reasonable to infer that these individuals received structured swimming education from their elementary school years or even earlier—and that this instruction was invariably provided by private swimming facilities.

\section{Conclusion}
This study evaluated nationwide spatial patterns of sports facility distribution in Japan, establishing a clear structural divergence: private sports facilities exhibit dense spatial clustering in high-radiance urban centers, whereas public facilities are distributed evenly across regions to promote spatial equity. Crucially, satellite nighttime light (NTL) radiance correlates strongly with private facility density, concentrating commercial fitness infrastructure near transit hubs, central business districts, and urban residential zones.

This facility distribution exerts a direct downstream effect on athletic development and competition outcomes. As illustrated by junior high and high school competitive swimming data (\figref{fig:swim_juniorhigh2021} and \figref{fig:swim_high2021}), prefectures characterized by higher NTL intensity produce a disproportionately larger share of national tournament qualifiers. Because swimming facilities entail high capital and maintenance costs, commercial venues are heavily concentrated in highly lit metropolitan centers (e.g., Tokyo, Kanagawa, Saitama, Osaka, Aichi). Consequently, youth athletes in low-radiance rural areas face structural accessibility barriers during early career formation—revealing a distinct form of "environmental disparity" in sports access that was previously overlooked in facility planning.

To expand upon these empirical insights, future work will integrate supplementary Earth observation datasets provided by JAXA's G-Portal \cite{gportal}. Specifically, I aim to incorporate environmental parameters from "SHIZUKU" (GCOM-W) \cite{gcom_w} (e.g., snowfall and precipitation metrics) to examine location dynamics in winter sports disciplines. Furthermore, vegetation metrics (NDVI/EVI) derived from "SHIKISAI" (GCOM-C) \cite{gcom_c} will be applied to investigate spatial disparities in outdoor sports infrastructure, urban green spaces, and localized cultural capital.

\paragraph{\texorpdfstring{Declaration of Generative AI in Scientific Writing}{Declaration of Generative AI in Scientific Writing}}
During the preparation of this work, the authors used ChatGPT-4 (OpenAI, accessed July 8, 2026) and Google Gemini (Google, accessed July 8, 2026) to enhance language clarity and readability. Specifically, these tools were used for \LaTeX{} formatting (including figure embedding and placement, table creation, and Bib\TeX{} reference adjustments), as well as for the translation and proofreading of English terms (such as grade level representations).

\bibliographystyle{plainnat}
\bibliography{jssij2024_fixed_en}

\end{document}